\documentclass[epj]{svjour}
\usepackage{psfig}
\usepackage{times}

\begin{document}

\title{On the possibility of $f_0$ observation in low energy
$pp$ collisions\thanks{\emph{Supported by Forschungszentrum J\"ulich,
BMBF  and DFG}}}
\author{E.L. Bratkovskaya$^1$, W. Cassing$^1$, L.A. Kondratyuk$^2$
and  A. Sibirtsev$^1$ \\}
\institute{$^1$ Institute for Theoretical Physics, \\
University of Giessen, D-35392 Giessen, Germany \\
$^2$ Institute of Theoretical and Experimental Physics, \\
117259 Moscow, Russia\\}

\date{Received: date / Revised version: date}
\abstract{Within the meson-exchange model we calculate $f_0$-meson
production cross section in ${\pi}N$ and $NN$ reactions and investigate
the possibility for $f_0$ observation via the $K\bar{K}$ decay mode in
$pp$ collisions.  Our studies indicate that an extraction of the $f_0$
signal is unlikely due to the large background from other
reaction channels.}

\PACS{{13.75.Cs} {Nucleon-nucleon interactions} \and
{13.75.Gx} {Pion-baryon interactions} }

\authorrunning{E.L. Bratkovskaya et al.}
\titlerunning{On possibility for $f_0$ observation in low energy
$pp$ collisions}
\maketitle

\section{Introduction}
The status of the scalar $f_0$-meson is still an
open problem in particle physics. In the 1996 Review of
Particle Physics~\cite{PDG1} the $f_0$ decay modes were announced
as $78.1\pm 2.4$\% for the $\pi\pi$ and $21.9\pm 2.4$\% for the
$K\bar{K}$-channel. In the 1998 Review~\cite{PDG2}, however,
the $f_0\to\pi\pi$ mode is established as $dominant$, while the
$f_0\to K\bar{K}$ mode is stated as $seen$.

A recent theoretical status of the problem has been presented by Oller
and Oset~\cite{Oller} and Krehl, Rapp and Speth~\cite{Krehl}. Here we
do not attempt to add a further summary on the problem, but discuss the
possibility for a direct observation of $f_0$-mesons in low energy
proton-proton collisions.  Our study is relevant to the  measurement of
$K^+K^-$ spectra from $pp$ interactions performed recently by the DISTO
Collaboration at SATURNE~\cite{DISTO} as well as to the current
experimental program at COSY~\cite{COSY}.

It is well known that the resonance spectral function is distorted if
one of the resonance decay channels has a threshold within the
resonance width~\cite{Bashinskii1}. A classical example is the scalar
$f_0$-meson with a pole mass slightly below the $K\bar{K}$ threshold,
but due to the finite $f_0$ width this decay channel is kinematically
allowed.  This leads to an enhanced $K\bar{K}$ production close to the
two kaon mass and was observed experimentally in $\pi N$
reactions~\cite{Dahl,Beusch,Pawlicki,Cason,Cohen,Etkin,Anisovich}.
Moreover, as was proposed by Bashinskii and
Kerbikov~\cite{Bashinskii2}, similar phenomena can be directly observed
in the $pd\to ^3HeX$ reaction close to the $K\bar{K}$ threshold.

Here we focus on the $K\bar{K}$ production in $pp$ reactions.  We start
with the paradigm proposed by Morgan and
Pennington~\cite{Morgan1,Morgan2} and use the  Breit-Wigner resonance
prescription for the $f_0$-meson, though keeping in mind the simplicity
of the BW approximation as e.g. pointed out by Janssen et
al.~\cite{Janssen}.

\section{The reaction $\pi N\to f_0 N$}
\begin{figure}[h]
\psfig{file=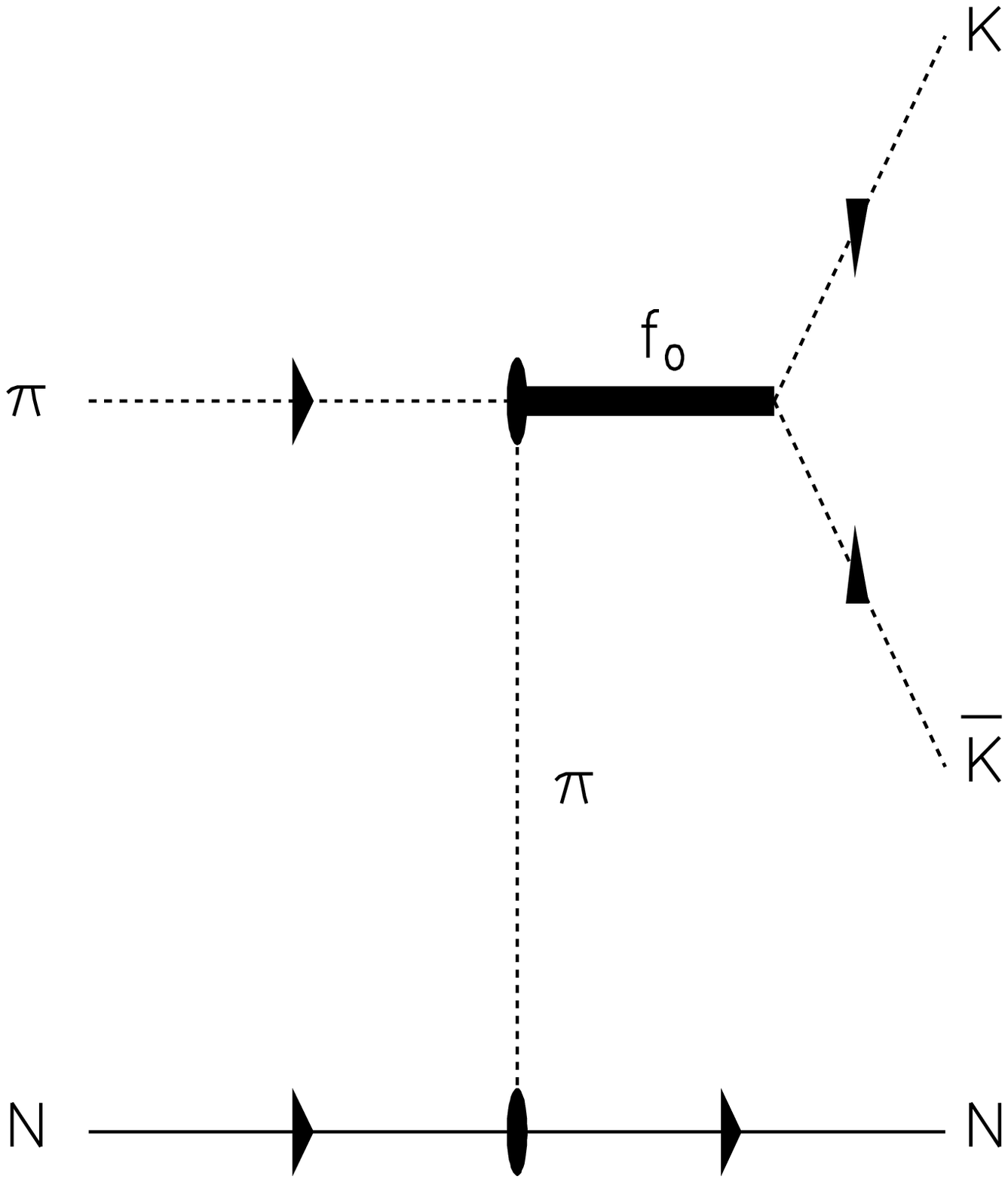,width=9cm,height=5cm}
\caption{Diagram for the ${\pi}N{\to}f_0N{\to}K\bar{K}N$
reaction.}
\label{Fig1}
\end{figure}

In order to test the validity of the approach~\cite{Morgan1,Morgan2}
we start with the ${\pi}N{\to}f_0N{\to}K\bar{K}N$ reaction.
The relevant one-pion exchange diagram is shown in
Fig.~\ref{Fig1}; the corresponding
differential cross section can be calculated as
\begin{eqnarray}
{d\sigma\over dM} =\int\limits_{t_{min}}^{t_{max}} dt \
{1\over 2^8\pi^3 \ s} \ {|{\bf k}| \over |{\bf q}|^2} |M_{if}|^2,
\label{dsdmpin}
\end{eqnarray}
where $M$ is the invariant $K\bar{K}$-mass, $s$ is the total energy of
the pion-nucleon system squared, ${\bf q}$ is the pion three-momentum
in the $\pi N$ center-of-mass frame, while ${\bf k}$ is the kaon
three-momentum in the $f_0$ rest frame and  $|{\bf k}|= \sqrt{M^2-4
m_K^2}/2$. In~(\ref{dsdmpin}) $t$ stands for the transfered
four-momentum squared $t=(p_2^\prime -p_2)^2$, where $p_2, p_2^\prime$
are the four-momenta of the nucleons in the initial and final states.

The matrix element in Eq.~(\ref{dsdmpin}) is given by
\begin{eqnarray}
M_{if} &=&   g_{\pi NN} \ \bar u(p_2^\prime) \ i \ \gamma_5 u(p_2)
\ { A_{\pi\pi\to KK}(M) \over t-m_\pi^2} \nonumber \\
&\times & F_{\pi NN}(t) \ F_{f_0\pi\pi}(t) .
\label{mtrpiN}
\end{eqnarray}
In principle, the $\pi\pi{\to}K\bar{K}$ amplitude
can be taken as a $K$-matrix
solution from the coupled channel analysis of the experimental
data on $\pi p$ and $\bar p p$ reactions
(cf.  Anisovich et al.~\cite{Anisovich}).
Another way is to adopt  the Breit-Wigner approach
and to define the $\pi\pi{\to}K\bar{K}$ amplitude  as
\begin{eqnarray}
A_{\pi\pi{\to}KK}(M) = {g_{f_0\pi\pi} \ g_{f_0KK} \over
M^2-m_{f_0}^2 + i \, m_{f_0}\Gamma_{tot}(M)},
\label{ApiK}
\end{eqnarray}
where $g_{f_0\pi\pi}$ and  $g_{f_0KK}$ denote the coupling constants,
while $m_{f_0}$ and $\Gamma_{tot}$ are the mass and width
of the $f_0$-meson, respectively.
Finally, the squared matrix element -- averaged over the initial and
summed over the final states -- is given as
\begin{eqnarray}
|M_{if}|^2 &=& {g_{\pi NN}^2 \ g_{f_0\pi\pi}^2 \ g_{f_0KK}^2  \over
(M^2 -m_{f_0}^2)^2 + m_{f_0}^2\Gamma_{tot}^2(M)}
\nonumber \\ &\times &
{-t\over (t-m_\pi^2)^2} \ \ F^2_{{\pi}NN}(t) \ F^2_{f_0\pi\pi}(t),
\label{matrpiNf}
\end{eqnarray}
where $F_{{\pi}NN}$ is the form factor at the ${\pi}NN$ vertex
taken in the monopole form
\begin{eqnarray}
F(t) = {\Lambda^2-m_\pi^2\over \Lambda^2-t}
\label{form}
\end{eqnarray}
with a cut-off parameter $\Lambda=1.05$~GeV~\cite{Tsushima}.
The ${\pi}NN$ coupling constant is
$g_{{\pi}NN}^2/4\pi =14.4$~\cite{Bonn}. $F_{f_0\pi\pi}$
is the form factor at the $f_0\pi\pi$ vertex taken
as in ~(\ref{form}) with $\Lambda=1.05$~GeV again.
Note that~(\ref{matrpiNf}) is valid only for low pion
energies, since at high energies one needs to Reggeize the
reaction amplitude similar to~\cite{Anisovich,Achasov1}.
Since the ${\pi}N \to f_0N\to K\bar{K}N$ cross section
depends upon the product  $g^2_{f_0\pi\pi} \cdot $
$g^2_{f_0KK}$ of the squared couplings and not their values itself,
one can fit only the product of the coupling constants by
experimental data.

The dominant $f_0$-meson decay channels  are the pion and
kaon modes~\cite{PDG2}. Neglecting  other possible
modes with extremely small decay branching ratios $Br$, one
has to saturate the unitarity condition:
\begin{eqnarray}
Br(f_0{\to}\pi\pi) + Br(f_0{\to}K\bar{K}) = 1.
\label{brkpi}
\end{eqnarray}
The branching ratios $Br(f_0\to \pi\pi)$ and  $Br(f_0\to K\bar{K})$
are given by integrals of the Breit-Wigner distribution over the
invariant mass of the final particles~\cite{Achasov95}:
\begin{eqnarray}
&&Br(f_0{\to}\pi\pi)=\!\!\int\limits_{2m_\pi}^\infty
\!{2dM\over \pi} {M \ m_{f_0} \ \Gamma_{f_0\pi\pi}(M)
\over (M^2-m_{f_0}^2)^2 + m_{f_0}^2 \Gamma_{tot}^2(M)}
\label{BrBW} \\
&&Br(f_0{\to}K\bar{K})\!=\!\!\int\limits_{2m_K}^\infty\!\!{2dM\over \pi}
{M \ m_{f_0} \ \Gamma_{f_0KK}(M)
\over (M^2-m_{f_0}^2)^2 + m_{f_0}^2 \Gamma_{tot}^2(M)},
\nonumber
\end{eqnarray}
where the total width of the $f_0$-meson  is defined as
\begin{eqnarray}
\Gamma_{tot}(M)= \left\{
\begin{array}{l}
\Gamma_{f_0\pi\pi}(M),\  \ \  \mbox{if}\  M \leq 2m_K\,, \\
\Gamma_{f_0\pi\pi}(M)+\Gamma_{f_0KK}(M),\ \ \ \mbox{if}\  M \geq 2m_K ,
\end{array} \right.
\label{sm31}
\end{eqnarray}
and the partial decay widths $f_0{\to}\pi\pi$ and $f_0{\to}K\bar{K}$ are
related to the relevant coupling constants as
\begin{eqnarray}
&&\Gamma_{f_0\pi\pi}(M) = {g_{f_0\pi\pi}^2\over 16\pi}
{\sqrt{M^2-4m_\pi^2} \over M^2}, \label{widpr}\\
&&\Gamma_{f_0KK}(M) = {g_{f_0KK}^2\over 16\pi}
{\sqrt{M^2-4m_K^2} \over M^2}. \nonumber
\end{eqnarray}

\begin{figure}[b]
\psfig{file=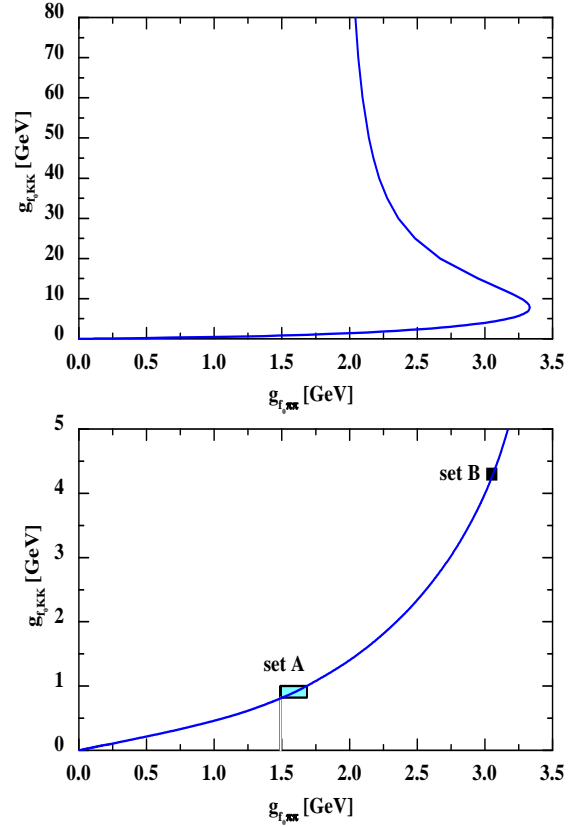,width=9cm,height=15cm}
\vspace*{-4cm}
\caption{Dependence of the coupling constants $g_{f_0K\bar{K}}$ on
$g_{f_0\pi\pi}$ according to the unitarity condition
(\protect\ref{brkpi}). $Set \ \ B$ denotes the constants
evaluated with the $f_0$ properties from the 1996 Review
of Particle Physics~\protect\cite{PDG1}. $Set \ A$
is our present estimation (see text).}
\label{Fig2}
\end{figure}

Substituting (\ref{BrBW}), (\ref{sm31}) and (\ref{widpr}) into
(\ref{brkpi}) one finds that  formula~(\ref{brkpi}) provides a unique
relation between the coupling constants $g_{f_0\pi\pi}$ and $g_{f_0K\bar{K}}$.
Fig.~\ref{Fig2} shows the result of our numerical solution
of~(\ref{brkpi}).  The upper part of Fig.~\ref{Fig2} displays
$g_{f_0KK}$ as a function of $g_{f_0\pi\pi}$ in a wide range. The
maximum value of $g_{f_0\pi\pi}$ is found to be $\simeq$3.33 and it
approaches an asymptotic  value of~1.93, whereas $g_{f_0KK}$ always
increases with $g_{f_0\pi\pi}$ for values below 3.33.  The lower part
of Fig.~\ref{Fig2} shows the latter range for $g_{f_0\pi\pi}$ and
$g_{f_0KK}$ on a larger scale.

In order to fix the $f_0\pi\pi$ and $f_0K\bar{K}$ coupling constants
individually one needs the explicit knowledge of one of the branching
ratios.  For instance, taking $Br(f_0{\to}\pi\pi)=$78.1\% \cite{PDG1}
we obtain $g_{f_0\pi\pi}$=3.05~GeV and $g_{f_0KK}$=4.3~GeV; according
to~(\ref{widpr}) and (\ref{sm31}) the total $f_0$ width then amounts to
233~MeV. Fig.~\ref{Fig2} shows this solution as $set \ \ B$.

\begin{figure}[t]
\psfig{file=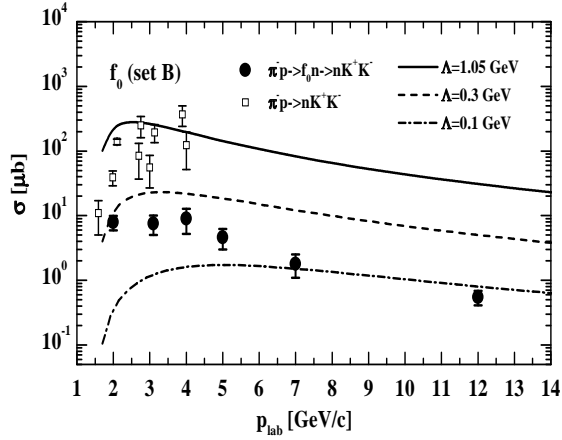,width=9cm,height=16cm}
\vspace*{-9.5cm}
\caption{The $\pi^-p{\to}f_0n{\to}K^+K^-n$
and $\pi^-p{\to}K^+K^-n$ cross sections. The lines show
calculations with the coupling constants from $set \ \ B$
for different cut-off parameters at the $f_0\pi\pi$
vertex. The experimental data are from Refs.~\protect\cite{Dahl,Beusch,LB}.}
\label{Fig3}
\end{figure}

The $\pi^-p{\to}f_0n{\to}K^+K^-n$ total cross section
calculated with $set \ \ B$ is shown by the solid line in
Fig.~\ref{Fig3} and  overestimate the experimental
data collected in Refs.~\cite{Dahl,Beusch}. Fig.~\ref{Fig3} also
shows the data~\cite{LB} for the $\pi^-p{\to}K^+K^-n$
cross section, which is substantially above the
$\pi^-p{\to}f_0n$ data since it includes other
production mechanisms as e.g. proposed in Ref.~\cite{Sibirtsev1}.

In principle, to fit the $\pi^-p{\to}f_0n{\to}K^+K^-n$
data with the coupling constants from $set \ \ B$
one might adjust the cut-off $\Lambda$
in ~(\ref{form}) at the $f_0\pi\pi$ vertex as a
free parameter. Fig.~\ref{Fig3} shows the calculations with
$\Lambda$=0.3~GeV and $\Lambda$=0.1~GeV, which are roughly
in line with the absolute magnitude for the
$\pi^-p{\to}f_0n$ cross section but contradict
its energy dependence.

We conclude that it is not possible to describe the
experimental data on the $\pi^-p{\to}f_0n{\to}K^+K^-n$
reaction adopting the $f_0\pi\pi$ and $f_0K\bar{K}$ coupling
constants from $set \ \ B$. Moreover, as was shown
above, $set \ \ B$ yields a large
total width for the $f_0$-meson (233~MeV) that is
out of the range $\Gamma_{f_0}$=40-100~MeV
quoted in the 1996-estimation from the Particle Data Group~\cite{PDG1}.

\begin{figure}
\psfig{file=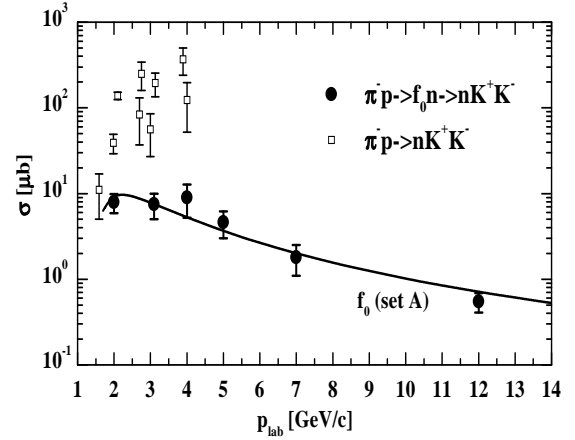,width=9cm,height=16cm}
\vspace*{-9.5cm}
\caption{The $\pi^-p{\to}f_0n{\to}K^+K^-n$ (full dots)
and $\pi^-p{\to}nK^+K^-$ (open squares) cross sections from
Refs.~\protect\cite{Dahl,Beusch,LB}. The solid line shows the
calculation with the coupling constants from $set \ \ A$.}
\label{Fig4}
\end{figure}

We now fit the data~\cite{Dahl,Beusch} on
the $\pi^-p{\to}f_0n{\to}K^+K^-n$ cross section by taking
the product of the $g_{f_0\pi\pi}$ and $g_{f_0KK}$ coupling
constants as a free parameter. The result is shown in
Fig.~\ref{Fig4}. Our solution for the product of the coupling
constants is shown in Fig.~\ref{Fig2} as $set \ \ A$
($g_{f_0\pi\pi}$=1.49~GeV, $g_{f_0KK}$=0.82~GeV)
that leads to the following $f_0$-meson
properties:
\begin{eqnarray}
&&Br(f_0{\to}\pi\pi)=98{\rm \%} \nonumber \\
&&Br(f_0{\to}K\bar{K})=2{\rm \%} \nonumber \\
&&\Gamma_{tot}=44.3~{\rm MeV},
\end{eqnarray}
which are in a nice agreement with the numbers from the recent
Review of Particle Physics~\cite{PDG2}.

\begin{figure}
\psfig{file=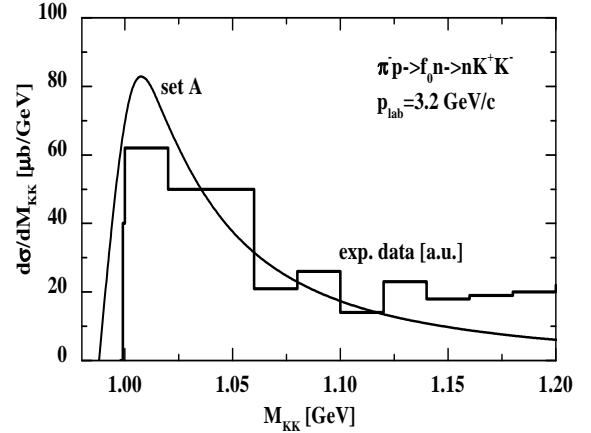,width=9cm,height=16cm}
\vspace*{-9.7cm}
\caption{The $K\bar{K}$ invariant mass distribution from
$\pi^-p$ reactions at 3.2~GeV/c. The histogram shows the
experimental data from Ref.~\protect\cite{Dahl}
while the solid line is our calculation with $set \ \ A$.}
\label{Fig5}
\end{figure}

Fig.~\ref{Fig5} displays the $K\bar{K}$ invariant mass spectrum from 
$\pi^-p$ collisions at a beam momentum of 3.2~GeV/c in comparison
to the experimental data from~\cite{Dahl}.
The solid line in Fig.~\ref{Fig5} indicates our calculation
with the parameters from $set \ \ A$ which
reasonably describes the experimental spectrum.

\section{The reaction $NN{\to}f_0NN$}
\begin{figure}[t]
\psfig{file=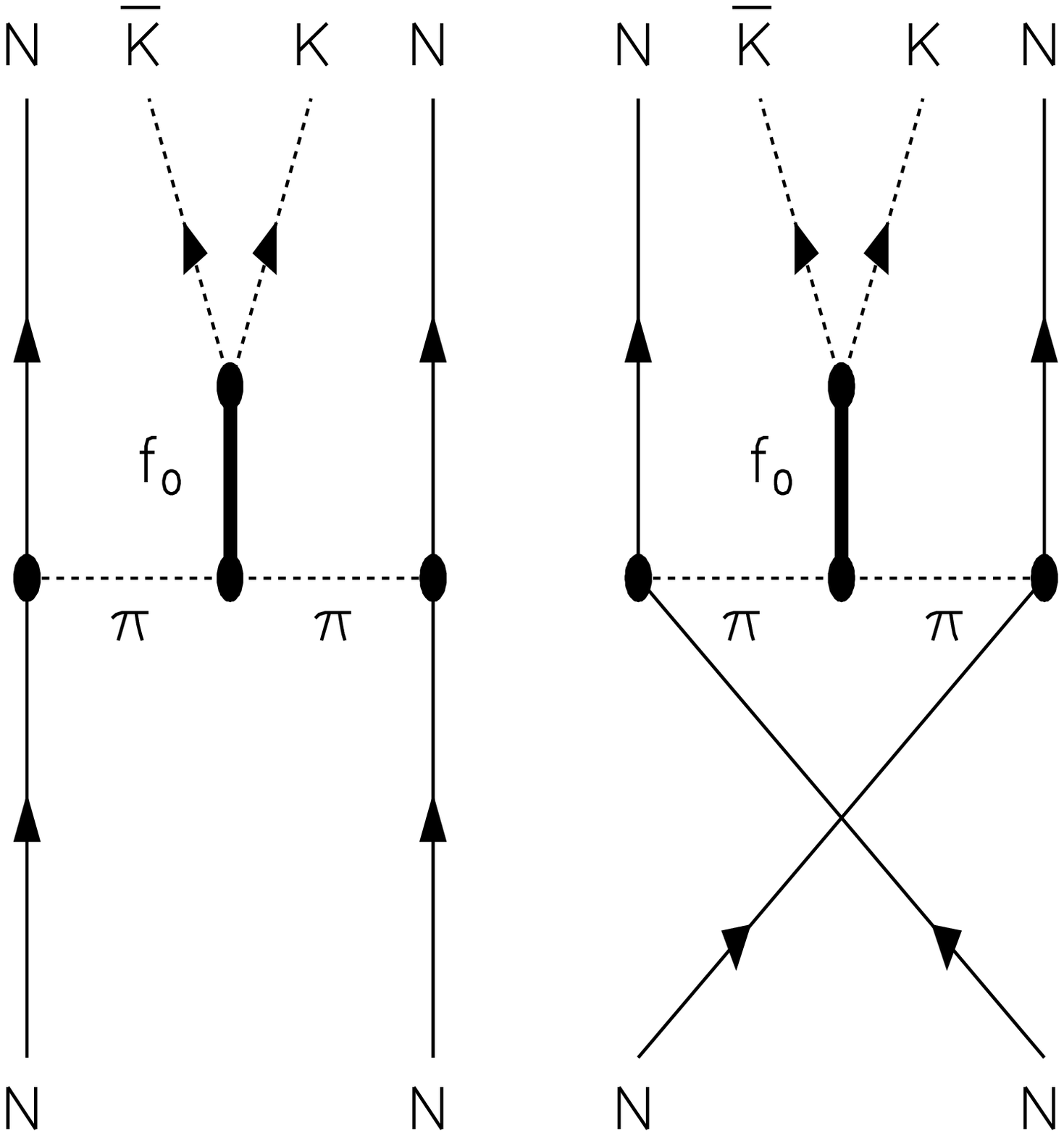,width=9cm,height=5cm}
\caption{The diagrams for the $NN{\to}f_0NN{\to}K\bar{K}NN$
reaction.}
\label{Fig6}
\end{figure}

The relevant diagrams for the $NN{\to}f_0NN{\to}K\bar{K}NN$
reaction are shown in Fig.~\ref{Fig6}; the corresponding
differential cross section is given as
\begin{eqnarray}
{d\sigma\over dM}\!=\!\!{\int}dE_1^\prime  dq_0  d{\cos}\theta_q
\, d\varphi_q \ {1\over 2^{11}\pi^6\sqrt{s}}
\, {|{\bf k}| \over |{\bf p}_1| }
\, |M_{if}|^2,
\label{dsdmpp}
\end{eqnarray}
with the matrix element taken as the sum of the direct and exchange
terms in Fig.~\ref{Fig6},
\begin{eqnarray}
M_{if}&=& g_{\pi NN}^2 \ \bar u(p_1^\prime)\ i \ \gamma_5 \ u(p_1) \
{ A_{\pi\pi\to KK}(M) \over (q_1^2-m_\pi^2) (q_2^2-m_\pi^2)}
\nonumber \\
&\times & \bar u(p_2^\prime) \ i \ \gamma_5 \ u(p_2)
\ F_{\pi NN}(q_1^2)\ F_{\pi NN}(q_2^2)
\nonumber \\
&-& g_{\pi NN}^2 \ \bar u(p_1^\prime) \ i \ \gamma_5 \ u(p_2) \
{ A_{\pi\pi\to KK}(M) \over (\tilde q_1^2-m_\pi^2)
(\tilde q_2^2-m_\pi^2)} \nonumber \\ &\times &
\bar u(p_2^\prime) \ i \ \gamma_5 \ u(p_1)
\ F_{\pi NN}(\tilde q_1^2)\ F_{\pi NN}(\tilde q_2^2),
\label{mtrNN}
\end{eqnarray}
where ${\bf k}$ is the kaon three-momentum in the $f_0$-rest frame,
$p_1$ and $p_2$ are the four-momenta of the initial nucleons,
while $p_1^\prime$ and $p_2^\prime$ are the four-momenta of the
final nucleons. Moreover, ${\bf p}_1$ is the
three-momentum of the initial nucleon in their center-of-mass frame
(cms), $E_1^\prime$ is the energy of the final nucleon in
the cms, $q_0$ and ${\bf q}$ are the energy and three-momentum of the kaon
pair in the cms, respectively. In~(\ref{mtrNN})
$\theta_q$ is the polar angle of the vector ${\bf q}$ in the cms
defined as $\theta_q=\widehat{{\bf q},{\bf p_1}}$,
while $\varphi_q$ is the
azimuthal angle of ${\bf q}$ in the cms.
The transfered 4-momenta are defined as:
$q_1= p_1^\prime - p_1$, $q_2= p_2-p_2^\prime$,
$\tilde q_1= p_1^\prime - p_2$ and $\tilde q_2 = p_1 - p_2^\prime$,
while $M$ denotes the invariant mass of the $K\bar K$-system.

The square of the matrix element (\ref{mtrNN}) -- averaged over the initial
and summed over the final states -- is given by
\begin{eqnarray}
|M_{if}|^2&=&\!\!{g_{\pi NN}^4 \, g_{f_0\pi\pi}^2 \, g_{f_0KK}^2 \over
(M^2-m_{f_0}^2)^2+m_{f_0}^2\Gamma_{tot}^2(M)}
\nonumber \\ &\times &
F_{\pi NN}^2(q_1^2) \, F_{\pi NN}^2(q_2^2)
{q_1^2 \ q_2^2 \over (q_1^2-m_\pi^2)^2(q_2^2-m_\pi^2)^2}
\nonumber\\
&+& {g_{\pi NN}^4 \, g_{f_0\pi\pi}^2 \, g_{f_0KK}^2 \over
(M^2-m_{f_0}^2)^2+m_{f_0}^2\Gamma_{tot}^2(M)}
\nonumber \\
&\times &
F_{\pi NN}^2(\tilde q_1^2) \, F_{\pi NN}^2(\tilde q_2^2)
{\tilde q_1^2 \ \tilde q_2^2 \over (\tilde q_1^2-m_\pi^2)^2
(\tilde q_2^2-m_\pi^2)^2} \nonumber\\
&+& \ { interference \ \ term}
\label{matrNNf}
\end{eqnarray}

Actually one has to introduce a form factor at the $f_0\pi\pi$
vertex since both pions are off their mass-shell. Following the
assumption from Refs.~\cite{Chung,Nakayama} we use the form
\begin{equation}
F_{f_0\pi\pi}(q_1^2,q_2^2)=F_{\pi NN}(q_1^2)F_{\pi NN}(q_2^2) ,
\label{form1}
\end{equation}
where the ${\pi}NN$ form factor was taken as in~(\ref{form})
with a cut-off parameter $\Lambda$=1.05~GeV. The form
factor~(\ref{form1}) is normalized to unity at
$q_1^2=m_\pi^2$ and $q_2^2=m_\pi^2$, which is consistent with
the kinematical conditions for the determination of the
$f_0\pi\pi$ coupling constant.

\begin{figure}[h]
\psfig{file=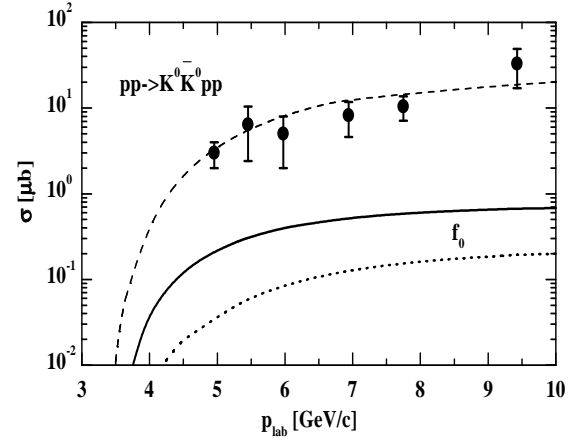,width=9cm,height=16cm}
\vspace*{-9.7cm}
\caption{The $pp{\to}f_0pp{\to}K^0\bar{K^0}pp$  cross section
calculated with coupling constants from $set \  A$ and with
(dotted line) and without form factor (solid line) at the
$f_0\pi\pi$ vertex. The experimental data for the
$pp{\to}K^0\bar{K^0}pp$ reaction are taken from Ref.~\protect\cite{LB},
while the dashed line shows the corresponding calculation within the
one-boson exchange model from Ref.~\protect\cite{Sibirtsev1}.}
\label{Fig7}
\end{figure}

The dotted line in Fig.~\ref{Fig7} shows the
$pp{\to}f_0pp{\to}K^0\bar{K^0}pp$ cross section calculated with the
coupling constants from $set \ \ A$ and with the form
factor~(\ref{form1}) in comparison to the experimental data~\cite{LB}
for the $pp{\to}K^0\bar{K^0}pp$ reaction. The dashed line shows the
calculations within the pion and kaon exchange model from
Ref.~\cite{Sibirtsev1} for $K\bar{K}$ production. To
estimate the maximal $f_0$ production cross section we neglect the form
factor at the $f_0\pi\pi$ vertex and show the result in terms of the
solid line in Fig.~\ref{Fig7}.  Actually, the contribution from $f_0$
production to the total $pp{\to}K^0\bar{K^0}pp$ cross section is almost
negligible at high energies. However, a possible way for $f_0$
observation is due to the low energy part of the $K\bar{K}$ invariant
mass spectrum.

\begin{figure}[b]
\psfig{file=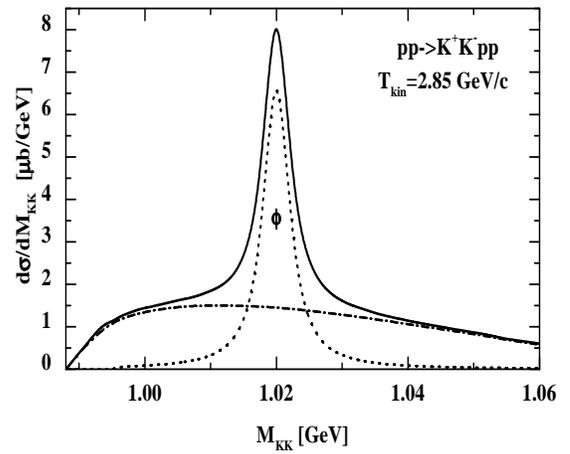,width=9cm,height=16cm}
\vspace*{-9.5cm}
\caption{The $K^+K^-$ invariant mass distribution from $pp$ collisions
at 2.85~GeV. The dash-dotted line shows the contribution from
$K^+K^-$-pair production according to the OBE model from
Ref.~\protect\cite{Sibirtsev1}; the dotted line is  the spectrum  from
the $pp{\to}{\phi}pp$ reaction while the solid line illustrates their sum.}
\label{Fig8}
\end{figure}
We thus calculate the $K^+K^-$ invariant mass spectrum from the
$pp{\to}K^+K^+pp$ reaction at a beam energy of 2.85 GeV, which
corresponds to the kinematical conditions for the DISTO experiment at
SATURNE \cite{DISTO}.  Since at this energy the $\phi$-meson production
becomes possible we include its contribution to the $K^+K^-$ spectrum.
The $pp{\to}{\phi}pp$ total cross section was taken from
Ref.~\cite{Sibirtsev2} and the $K^+K^-$ invariant mass was distributed
according to the Breit-Wigner resonance prescription with  a full
$\phi$-meson width $\Gamma_\phi$ = 4.43~MeV and the branching ratio
$Br(\phi\to K^+K^-)=49.1$\% \cite{PDG2}.

The dotted line in Fig.~\ref{Fig8} shows the $K^+K^-$ invariant mass
spectrum for the $pp{\to}{\phi}pp$ reaction while the dash-dotted line
indicates the spectrum from the $pp{\to}K^+K^-pp$ reaction, which was
calculated as in Ref.~\cite{Sibirtsev1} on the basis of pion and kaon
exchange diagrams.  The solid line in  Fig.~\ref{Fig8} shows the total
$K^+K^-$ spectrum.

\begin{figure}[t]
\psfig{file=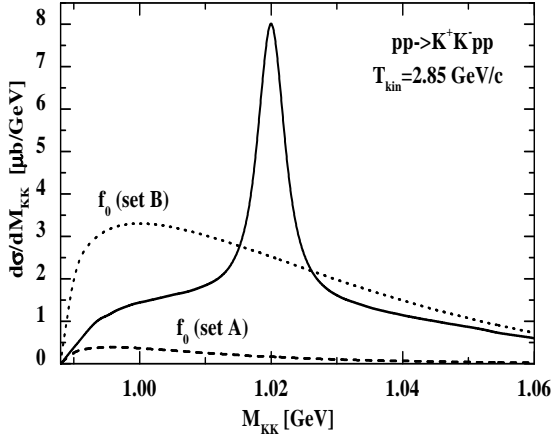,width=9cm,height=16cm}
\vspace*{-9.7cm}
\caption{The $K^+K^-$ invariant mass distribution from
$pp$ collisions at 2.85~GeV. The solid line is the same as in
Fig.~\protect\ref{Fig8}. The dashed line shows the
contribution from the $pp{\to}f_0pp{\to}K^+K^-pp$ reaction
calculated with constants from $set \ A$, while the
dotted line shows the result obtained with $set \ B$.}
\label{Fig9}
\end{figure}

To test the possibility for a direct $f_0$ observation via the $K^+K^-$
spectrum from $pp$ collisions one should compare the total $K^+K^-$
production cross section from meson-exchange diagrams and $\phi$-decay
(denoted as background) with the explicit contribution from the $pp\to
f_0pp\to K^+K^-pp$ reaction.  The solid line in Fig.~\ref{Fig9}  shows
the background while the dashed line indicates the $K^+K^-$ spectrum
calculated with the coupling constants from $set \ \ A$ and without
form factor at the $f_0\pi\pi$ vertex.  If the $f_0\pi\pi$ and
$f_0K\bar{K}$ coupling constants are determined by the $set \ \ A$,
than it is quite obvious that the $f_0$-meson cannot be directly
detected in $pp$ collisions by using the $K^+K^-$-mode. Note, that when
introducing a form factor~(\ref{form1}) at the $f_0\pi\pi$ vertex the
contribution from $pp{\to}f_0pp{\to}K^+K^-pp$ becomes even smaller.

To test the sensitivity of the model upon the $f_0$ parameters we also
perform the calculation with $set \ \ B$ and show the result in terms
of the dotted line in Fig.~\ref{Fig9}. Indeed, in that case the $f_0$
contribution is very strong at low $K^+K^-$ invariant mass. Thus
experimental data from DISTO might be crucial for the examination of
the $f_0$ properties.

\begin{figure}[h]
\psfig{file=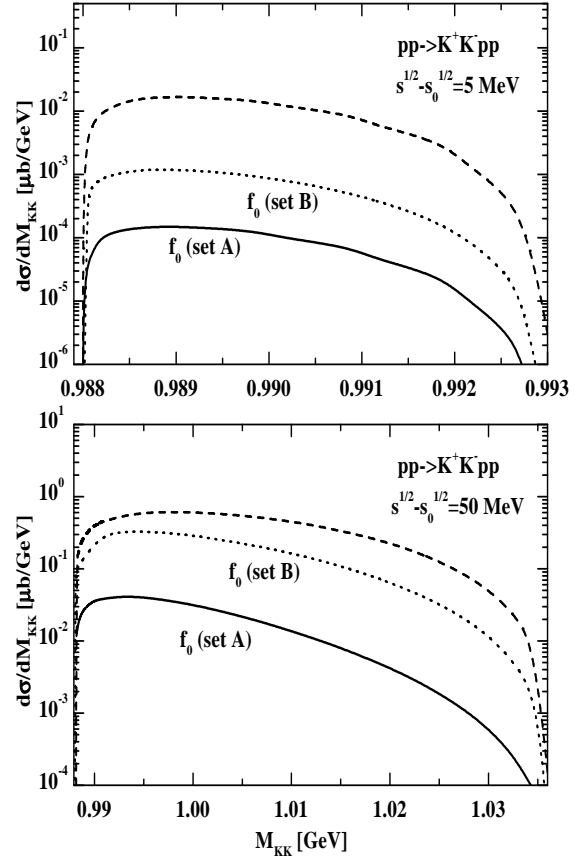,width=9cm,height=16cm}
\vspace*{-3.7cm}
\caption{The $K^+K^-$ invariant mass distribution from $pp$ collisions
at excess energies of 5 (upper part) and 50~MeV (lower part).  The
dashed lines indicate the contribution from the $pp{\to}K^+K^-pp$
reaction according to Ref.~\protect\cite{Sibirtsev1}.  The solid lines
show the $K^+K^-$ spectrum from the $pp{\to}f_0pp{\to}K^+K^-pp$
reaction calculated with the coupling constants from $set \ A$, while
the dotted lines are the calculations with $set \ B$.}
\label{Fig10}
\end{figure}

We also test the possibility for $f_0$ detection by use of the $K^+K^-$
spectrum from $pp$ collisions at energies very close to the
$pp{\to}K^+K^-pp$ reaction threshold, i.e. for the kinematical
conditions available at the COSY accelerator. Fig.~\ref{Fig10} shows
our calculations for the excess energies $\sqrt{s}-2m_N-2m_K$ of 5 and
50~MeV. The dashed lines in Fig.~\ref{Fig10} show the $K^+K^-$
production calculated again in accordance with~\cite{Sibirtsev1}.  The
solid lines correspond to  our results obtained with the $f_0\pi\pi$
and $f_0K\bar{K}$ coupling constants from $set \ \ A$, while the dotted
lines are calculations with $set \ \ B$.  Note that the results shown
in Fig.~\ref{Fig10} are obtained without a form factor at the
$f_0\pi\pi$ vertex, thus they should be considered as upper limits for
the $f_0$-meson contribution. Furthermore, at an excess energy of 5~MeV
strong final state interactions between the two protons should enhance
the yield substantially. However, these final state interactions are of
similar strength in both reaction channels and may be disregarded in
their ratio, which is the relevant quantity here.

We conclude that at an excess energy of 5~MeV, i.e. very close to the
$pp{\to}K^+K^-pp$ reaction threshold, the contribution from $f_0$-meson
production is almost negligible and cannot be separated from the
background processes.  At $\sqrt{s}-2m_N-2m_K$=50 MeV the contribution from 
the $pp{\to}f_0pp{\to}K^+K^-pp$ reaction, as a maximal estimation,
is a few times less  than the contribution from $K^+K^-$-pair
production due to pion and kaon exchange diagrams.

\section{Conclusions}
We have investigated the production of $f_0$-mesons in $pp$
interactions and the possibility for its observation via the
$f_0{\to}K\bar{K}$ mode. Our calculations have been based upon the
one-pion exchange model and a Breit-Wigner prescription for the $f_0$
resonance which allows for a quantitative estimate.  The coupling
constants at the $f_0\pi\pi$ and $f_0K\bar{K}$ verticies have been
constraint by experimental data on the ${\pi}N{\to}f_0N$ reaction; this
approach gives a full $f_0$-meson width of 44.3~MeV and the branching
ratios $Br(f_0{\to}\pi\pi)$ = 98\%, $Br(f_0{\to}K\bar{K})$ = 2\%. Our
estimation is in line with the $f_0$ properties from the recent 1998
Review of Particle Physics~\cite{PDG2}, but substantially contradicts
the numbers from the 1996-Review~\cite{PDG1}.

It is found that the $K^+K^-$ invariant mass distribution from the
$pp{\to}K^+K^-pp$ reaction at a beam energy of 2.85~GeV, which
corresponds to the experimental condition for the DISTO experiment at
SATURNE, might be sensitive to the $f_0$-meson properties. With the
$f_0$ properties given by the 1996 Review of Particle
Physics~\cite{PDG1} the $f_0$ signal should be seen as an enhancement
in the low energy part of the $K^+K^-$-mass spectum. However, following
our estimation, we do not expect such an enhancement and predict a
$K^+K^-$ invariant mass spectrum as shown in Fig.~\ref{Fig9} ($set \ A$).

The possibility for the $f_0$-meson observation in $pp\to K\bar{K}pp$
reactions  at near-threshold energies available at the COSY accelerator
was studied, too.  Our calculations indicate that no $f_0$ signal might
be extracted from the $K^+K^-$ invariant mass spectrum due to the large
background contribution from other reaction
channels~\cite{Sibirtsev1} as arising from pion and kaon exchanges.

\begin{acknowledgement}
We appreciate   stimulating discussions with C. Hanhart,
J. Hai\-denbauer, W. K\"uhn, J. Ritman and M.A. Smondyrev.
\end{acknowledgement}

\end{document}